# Nonmagnetic Impurity Effects on Majorana Braiding Quantum Gates in Vortex Cores

Jiangwei Long [1], Zihui Liu [1], Yizhi Li [2] and Lijun Meng [1,†]

[1] School of Physics and Optoelectronics, Xiangtan University, Xiangtan 411105, Hunan, People's Republic of China

[2] School of Physics and Electronic Science, Hunan Institute of Science and Technology, Yueyang 414006, People's Republic of China

Using time-dependent many-body simulations, we investigate nonmagnetic impurity effects on vortex-core Majorana braiding. In the two-MZM $\sqrt{Z}$ gate, sufficiently strong attractive impurities suppress the topological gap $\Delta_t$, reduce adiabaticity, and degrade the quantization of the geometric-phase difference; the repulsive impurities considered here cause no comparable damage despite stronger zero-energy nonequilibrium local density of states (LDOS) distortions. Slowing the braid or using a $2[a_0]$ detour restores near-quantization in the cases examined. In the four-MZM configuration, attractive impurities leave the low-energy spectrum nearly unchanged, whereas repulsive impurities mainly suppress the first excited state. Slower braiding substantially improves the $X$ gate success probability. A specific attractive-impurity pair yields near-unit success (≈0.9985), unlike two weak repulsive impurities. These findings clarify how nonmagnetic impurities affect the low-energy spectrum and adiabatic dynamics of vortex-core Majorana braiding and provide theoretical guidance for designing robust braiding protocols in realistic topological superconducting platforms.

## I Introduction

Topological quantum computation (TQC) encodes quantum information nonlocally in anyonic quasiparticles and manipulates it through braiding and fusion, whose outcomes depend only on the topology of the anyon world lines and are therefore intrinsically robust against local noise and perturbations [1,2]. The theoretical framework of braid-group representations has been extensively developed [3,4], and the search for non-Abelian anyons in condensed-matter systems has become one of the central experimental frontiers.

Among the candidates, Majorana zero modes (MZMs)—zero-energy quasiparticles that are their own antiparticles and realize Ising anyons—are regarded as the most promising [5,6]. They were predicted to emerge at the ends of spin-orbit coupled semiconductor-superconductor nanowires [7], in the cores of vortices in topological superconductors [8], and in related heterostructures [9]. Over the past decade and a half, zero-bias conductance peaks consistent with MZMs have been reported in hybrid nanowires [10-13], ferromagnetic atomic chains on superconductors [14], and vortex cores of iron-based superconductors [15,16], complemented by observations of half-integer thermal transport in candidate topological phases [17]. Nevertheless, these signatures can often be mimicked by trivial in-gap states—such as Caroli–de Gennes–Matricon vortex levels [18] or impurity-induced Andreev bound states [19]—so that the unambiguous identification and controlled manipulation of MZMs remain outstanding challenges.

† Corresponding author. E-mail: ljmeng@xtu.edu.cn

A fundamental limitation arises from the Ising fusion and braiding rules themselves: exchange operations among MZMs generate only a discrete, nonuniversal subset of quantum gates—the Clifford group [20], so that universal quantum computation requires additional resources, such as magic-state distillation [21] and ancilla-based protocols [22], adiabatic manipulation schemes in extended networks [23], parity-assisted operations [24], or controlled couplings between Majorana modes [25,26]. Determining exactly which gates can be constructed through braiding alone, and how faithfully they can be realized in realistic materials, is therefore an essential step toward any experimental gate architecture.

Vortex-core MZMs in iron-based topological superconductors and magnetic-superconductor hybrid structures are particularly appealing in this context, since vortices can be created, displaced, and rearranged by external fields [27-29]. Recent time-dependent many-body simulations have shown that physically moving vortices along designed trajectories realizes elementary gates $\{\sqrt{Z}, \sqrt{X}, \text{and } X\}$, through the non-Abelian geometric phases accumulated during braiding [30,31], and that the real-space motion of vortex-core zero modes can be tracked via the time evolution of the nonequilibrium local density of states (LDOS) [32].

Real materials, however, unavoidably contain impurities and disorder, which are known to affect vortex-core electronic structure in profound ways: magnetic impurities create Yu-Shiba-Rusinov states and suppress the superconducting gap [19], whereas nonmagnetic impurities can enhance the visibility of vortex-core MZMs by pushing away the nearby Caroli–de Gennes–Matricon states counterintuitively [33]. On the dynamical side, nonadiabatic effects are known to limit braiding fidelity [34], and the influence of impurities, disorder, and interactions on braiding dynamics has begun to be explored [35,36]. A systematic understanding of how nonmagnetic impurities affect the construction of quantum gates by vortex-core MZM braiding is still lacking.

In this work, we investigate this problem using time-dependent many-body simulations of braiding-based gates constructed from vortex-core MZMs in the presence of nonmagnetic impurities. Both prototypical gate constructions are considered: the two-MZM configuration, in which a single exchange realizes the $\sqrt{Z}$ gate, and the four-MZM configuration, in which one and two exchanges realize the $\sqrt{X}$ and $X$ gates, respectively. The gate quality is quantified by the geometric-phase difference and by the transition probabilities between the relevant many-body states. For both configurations, we examine how the braiding dynamics depends on the impurity parameters and assess strategies for mitigating impurity-induced errors, including adjusting the braiding speed, rerouting the vortex trajectories, and, for the four-MZM configuration, introducing a second impurity. This paper is organized as follows: Sec. II describes the model and the simulation method; Sec. III presents the results for the two-MZMs $\sqrt{Z}$ gate and the four-MZMs $\sqrt{X}$ and $X$ gates; and Sec. IV summarizes our conclusions.

**II Theoretical Framework and Computational Methods**

Our starting point is the lattice model for vortex-core Majorana braiding, on a two-dimensional square lattice, an $s$-wave superconducting order parameter (SCOP), Rashba spin-orbit coupling (RSO), and exchange splitting jointly place the system in a class-D topological phase with Chern number $C = -1$,

$$\begin{aligned}\mathcal{H} = &-t_e \sum_{r,r',\beta} e^{i\theta_{rr'}} c^{\dagger}_{r,\beta} c_{r',\beta} - \mu \sum_{r,\beta} c^{\dagger}_{r,\beta} c_{r,\beta} + i\alpha \sum_{r,\delta,\beta,\gamma} e^{i\theta_{rr+\delta}} c^{\dagger}_{r,\beta} \left(\delta \times \sigma\right)^{z}_{\beta,\gamma} c_{r+\delta,\gamma} \\ &+ \sum_{r} \left(\Delta(r,t) c^{\dagger}_{r,\uparrow} c^{\dagger}_{r,\downarrow} + \Delta^{*}(r,t) c_{r,\downarrow} c_{r,\uparrow}\right) + JS \sum_{r} \left(c^{\dagger}_{r,\uparrow} c_{r,\uparrow} - c^{\dagger}_{r,\downarrow} c_{r,\downarrow}\right),\end{aligned} \tag{1}$$

where $-t_e$ is the nearest-neighbor hopping, $\mu$ is the chemical potential, $\alpha$ is the RSO strength, $JS$ is the ferromagnetic exchange splitting, and $\theta_{r,r'} = \frac{\pi n_{\phi}}{2N_x N_y}\left(xy' - yx'\right)$ is the Peierls phase ($n_{\phi}$ is the number of flux quanta in the supercell, $N_x$ ($N_y$) is size of square lattice, and r = $(x, y) \rightarrow$ r′ = $(x', y')$ the initial and final hopping sites. The gauge structure under periodic boundary conditions is given in SM Sec. I of Ref.[31]). This Hamiltonian describes the topological phase of $FeSe_{1-x}Te_x$ [37,38] and magnet-superconductor hybrid (MSH) systems [39,40].

We introduce nonmagnetic impurities in Eq. (1). An impurity at $r_0$ is described by a Gaussian potential:

$$V_i = V_0 \exp\left(-\frac{\left|r_i - r_0\right|^2}{2\eta^2}\right), H_{\text{imp}} = \sum_{i\sigma} V_i c^{\dagger}_{i\sigma} c_{i\sigma}, \tag{2}$$

with strength $V_0$ and spatial extent $\eta$. Since the potential is spin independent, it enters the BdG structure as $+V_i$ in the particle block and $-V_i$ in the hole block, therefore preserves both particle-hole symmetry, and fermion-parity conservation. We note that this potential serves as a vortex pinning center in Ref. [33]; here, by contrast, the impurity is static at $r_0$ while the vortex trajectories are prescribed by the gate protocol, which cleanly isolates the influence of the impurity on the braiding dynamics and gate operation.

The vortices move along prescribed trajectories; the magnitude and phase of the SCOP $\Delta(r)$, together with the smooth interpolation of the vortex positions between neighboring sites, follow Ref. [31] (Eqs. (2) and (3) and SM Sec. III). Adiabaticity requires $t_V \gg \hbar/\Delta_t$, where the topological gap $\Delta_t$ coincides with the energy of the lowest CdGM state, so as to avoid excitations between the vortex-core MZM and trivial CdGM states. All times are measured in units of the electronic time scale $\tau_e = \hbar/t_e$.

The initial and final states of the braid are fixed by $H(0)$. Spatially localized MZM single-particle states are obtained: diagonalizing $H(0)$ to extract the near-zero modes of the Majorana sector, performing a Takagi factorization [41,42] of the charge-conjugation operator and a subsequent localization with the position operator, and finally pairing the MZMs into fermionic modes. For four vortices, the two fermionic modes $d_1$, $d_2$ span the many-body Fock space, and the states $|00\rangle$, $|01\rangle$, $|10\rangle$, $|11\rangle$ are built from the quasiparticle vacuum [43]; for two vortices only a single fermionic mode is involved. The dynamics follows the time-dependent BdG equations, and many-body overlaps are evaluated with the Onishi formula [44]. For the technical details mentioned above, please refer to Ref. [31] (SM Sec. IV)

For the two-vortex $\sqrt{Z}$ gate we use the geometric phase difference as the diagnostic [45,46],

$$\Delta\phi^g = \phi_e^g - \phi_o^g, \tag{3}$$

where $\phi_i^g(t) = \arg\langle\Psi_i(t_0)|\Psi_i(t)\rangle - \mathrm{Im}\int_{t_0}^{t}\langle\Psi_i(t')|\partial_{t'}\Psi_i(t')\rangle dt'$, it is a gauge- and parametrization-invariant functional; Ising statistics requires $\Delta\phi^g$ to be an odd multiple of $\pi/2$ after a single exchange. Following Ref. [31], the geometric phase is far more sensitive to nonadiabatic effects than the fidelity, and we adopt it as the primary indicator for the $\sqrt{Z}$ gate.

The $\sqrt{X}$ and $X$ gate of the four-vortex system is executed in the odd-parity sector $\{|10\rangle, |01\rangle\}$, where the $|01\rangle - |10\rangle$ energy difference can be nearly reduced to zero, thereby suppressing spurious dynamical phases [30]. Starting from $|10(t=0)\rangle = |10\rangle$, the success and failure probabilities are defined as

$$p_{10}^{s}(t) = \left|\langle 01|10(t)\rangle\right|^2, \quad p_{10}^{f}(t) = \left|\langle 10|10(t)\rangle\right|^2. \tag{4}$$

A single exchange realizes the $\sqrt{X}$ gate (with $p_{10}^s = p_{10}^f = 1/2$ at $t = t_{1/2}$), and two exchanges realizes the $X$ gate (with $p_{10}^s = 1$, $p_{10}^f = 0$).

To visualize the gate operation in space and time, we employ the time-, energy-, and spatially resolved nonequilibrium LDOS:

$$N_{\mathrm{neq}}(t,\omega,\mathrm{r}) = (-1/\pi)\,\mathrm{Im}\left[G^r(t,\omega,\mathrm{r})\right], \tag{5}$$

where the retarded Green's function matrix follows from $\left[i\,\mathrm{d}/\mathrm{d}t + \omega + i\Gamma - \hat{H}(t)\right]\hat{G}^r(t,\omega) = \hat{1}$, with $\hbar/\Gamma$ the quasiparticle lifetime. $N_{\mathrm{neq}}$ has been shown to be proportional to the time-dependent differential conductance measured in STM experiments [32], and thus provides an experimentally accessible fingerprint of the braiding process: tracking its zero-energy component $N_{\rm neq}$ ($t$, $\omega$ = 0, $r$) reveals how the vortex-core MZMs move.

## III. Numerical Results

Parameters are taken from Ref. [31]. For the two-vortex $\sqrt{Z}$ gate ($N_x = N_y = 21$, i.e., a $21a_0 \times 21\,a_0$ lattice), the geometric phase difference is computed with ($\mu$, $\alpha$, $\Delta_0$, $JS$) = (−4, 1.6, 2.4, 3.09)$t_e$, whereas the corresponding $N_{\rm neq}$ maps are obtained with ($\mu$, $\alpha$, $\Delta_0$, $JS$) = (−7, 0.9, 2.4, 4.4) $t_e$. For the four-vortex $\sqrt{X}$ and $X$ gates ($N_x = N_y = 29$), both the braiding probabilities $p_{10}^s$, $p_{10}^f$ and the $N_{\rm neq}$ maps are computed with ($\mu$, $\alpha$, $\Delta_0$, $JS$) = (−7, 0.9, 2.4, 4.4) $t_e$. All calculations use $R_V = a_0$ and $\Gamma = 0.01t_e$, $\eta = 1$.

### A. Nonmagnetic impurity effects on $\sqrt{Z}$ Gate Construction

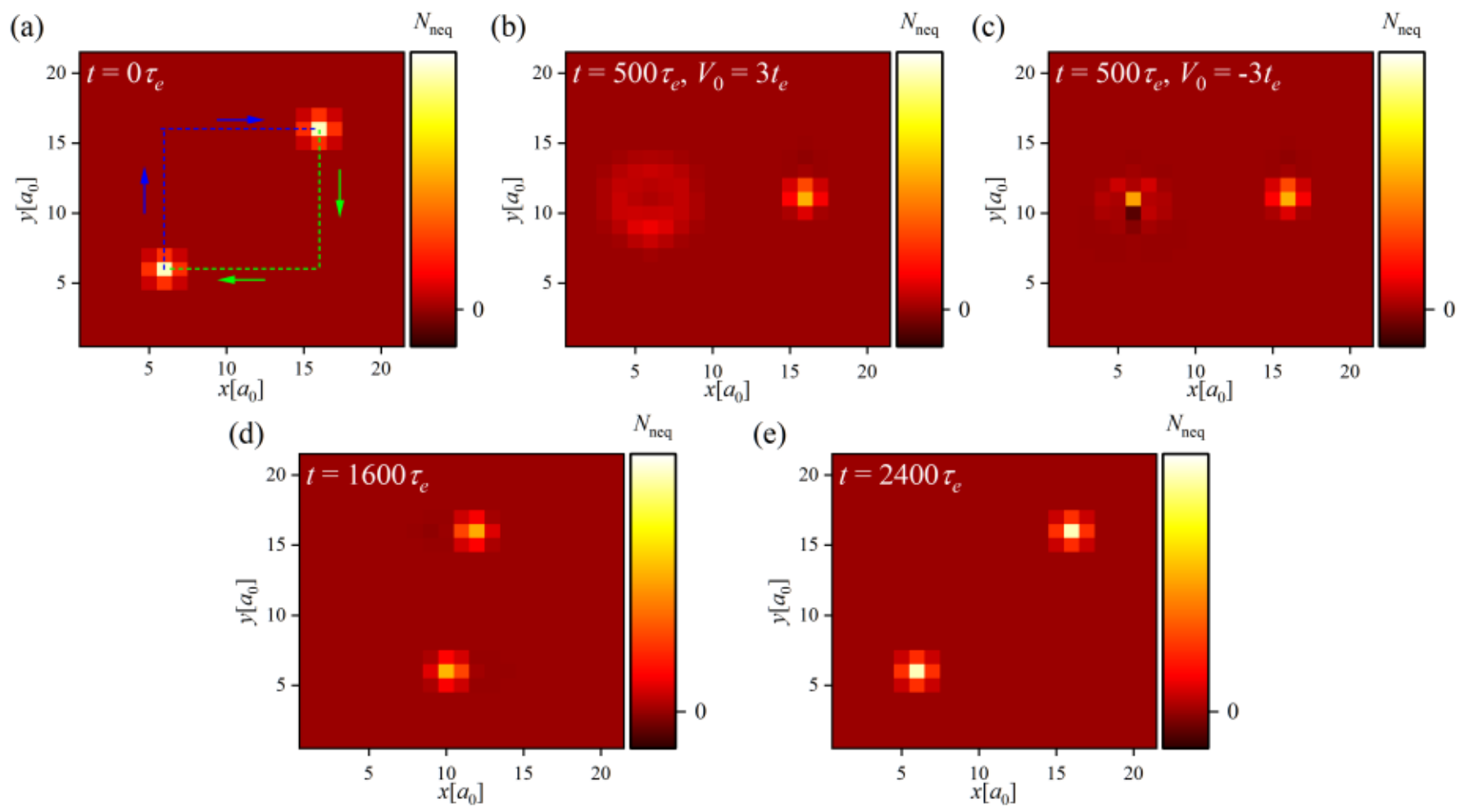

FIG. 1. Nonequilibrium zero-energy LDOS $N_{\rm neq}(t, \omega = 0, r)$ visualizing the braiding of two vortex-core MZMs for $\sqrt{Z}$ gate at (a) $t = 0\tau_e$, (b) $t = 500\tau_e$, $V_0 = 3t_e$, (c) $t = 500\tau_e$, $V_0 = -3t_e$, (d) $t = 1600\tau_e$, and (e) $t = 2400\tau_e$. The blue (green) arrow indicates the direction of motion of the two moving vortices initially located at (6, 6), (16, 16); dashed lines mark the vortex trajectories. Distances are in units of the lattice constant $a_0$. $t_v = 100\tau_e$.

The braiding of the MZMs is visualized using $N_{\rm neq}$ defined in Eq. (5). As shown in Fig. 1, the vortices move synchronously along the braiding path of Ref. [31] with $t_v = 100\tau_e$, while the nonmagnetic impurity of Eq. (2) is located at (6,11). The vortex initially at (6,6) crosses the impurity center at $t = 500\tau_e$. For the repulsive impurity ($V_0 = +3t_e$), the zero-energy weight at the impurity center is strongly suppressed and redistributed to surrounding sites, transforming the original sharp peak into a ring-like structure with a suppressed center [Fig. 1(b)]. For the attractive impurity ($V_0 = -3t_e$), the central peak remains, but the surrounding weight is strongly suppressed, yielding a narrower spike accompanied by a pronounced negative side lobe [Fig. 1(c)]. In both cases, the peak of the distant vortex is nearly unchanged. By $t = 1600\tau_e$, the $N_{\rm neq}$ maps have essentially recovered their clean form, and they remain close to the clean-system maps at $t = 2400\tau_e$ [Figs. 1(d) and 1(e)].

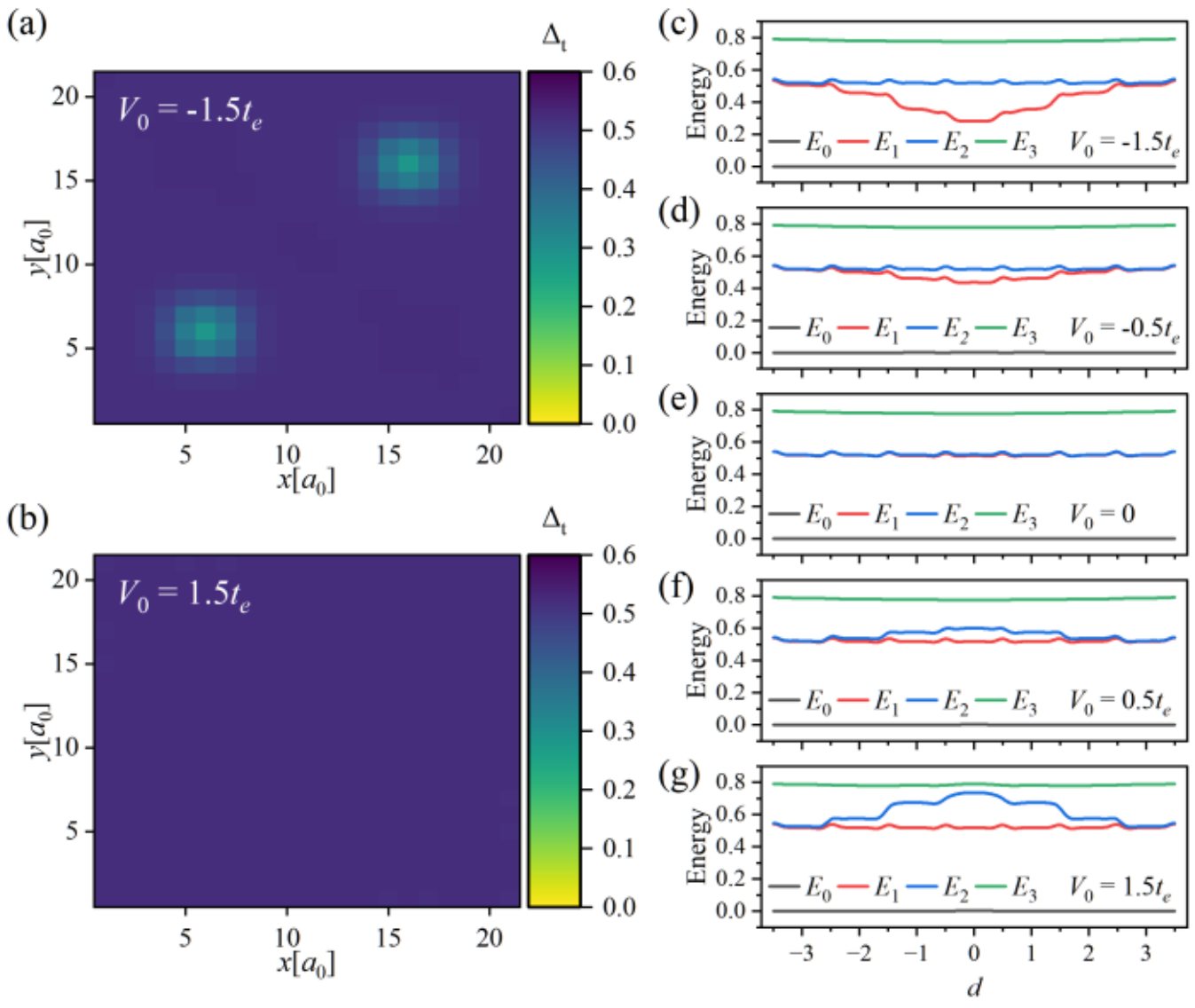

FIG. 2. Effect of a nonmagnetic impurity on the near-zero-energy spectrum of two-vortex configuration. (a), (b) Topological gap $\Delta_t$ (i.e., the energy of the first excited state) as a function of the impurity position for (a) $V_0 = -1.5t_e$ and (b) $V_0 = 1.5t_e$; (c)–(g) Energies of the MZM and of the first, second, and third excited states as a function of the vortex–impurity distance $d$ for (c) $V_0 = -1.5t_e$, (d) $V_0 = -0.5t_e$, (e) $V_0 = 0$, (f) $V_0 = 0.5t_e$, and (g) $V_0 = 1.5t_e$. Energies are in units of $t_e$. Two vortices initially located at (6, 6), (16, 16).

The low-energy spectrum of the two-vortex configuration depends strongly on both the impurity sign and the vortex–impurity distance as show in Fig. 2. Figs. 2(a) and 2(b) show the $\Delta_t$ as a function of impurity position for vortices initially located at (6,6) and (16,16). When the impurity is far from the vortex cores, $\Delta_t \simeq 0.52t_e$, essentially identical to the clean-system value. As a vortex crosses the impurity at (6,11), an attractive impurity pulls the overlapping vortex-core excitation toward zero energy. At $d = 0$, $\Delta_t \simeq 0.44t_e$ and $0.28t_e$ for $V_0 = -0.5t_e$ and $-1.5t_e$, respectively, whereas the second excited state remains near $0.52t_e$. By contrast, a repulsive impurity shifts the affected branch upward, while a nearly degenerate partner becomes the first excited state. Consequently, $\Delta_t$ closely follows the clean-system trajectory without a pronounced suppression. At $d = 0$, increasing $V_0$ from 0 to 1.5te raises the second excited state from $0.52t_e$ to $0.74t_e$, while the first excited state remains near $0.52t_e$. The gap bottleneck induced by attractive impurities may promote nonadiabatic transitions, whereas no comparable bottleneck occurs for the repulsive impurities considered here; whether the geometric phase is degraded or remains quantized is further determined from the geometric-phase difference in Eq. (3), calculated through time-dependent evolution.

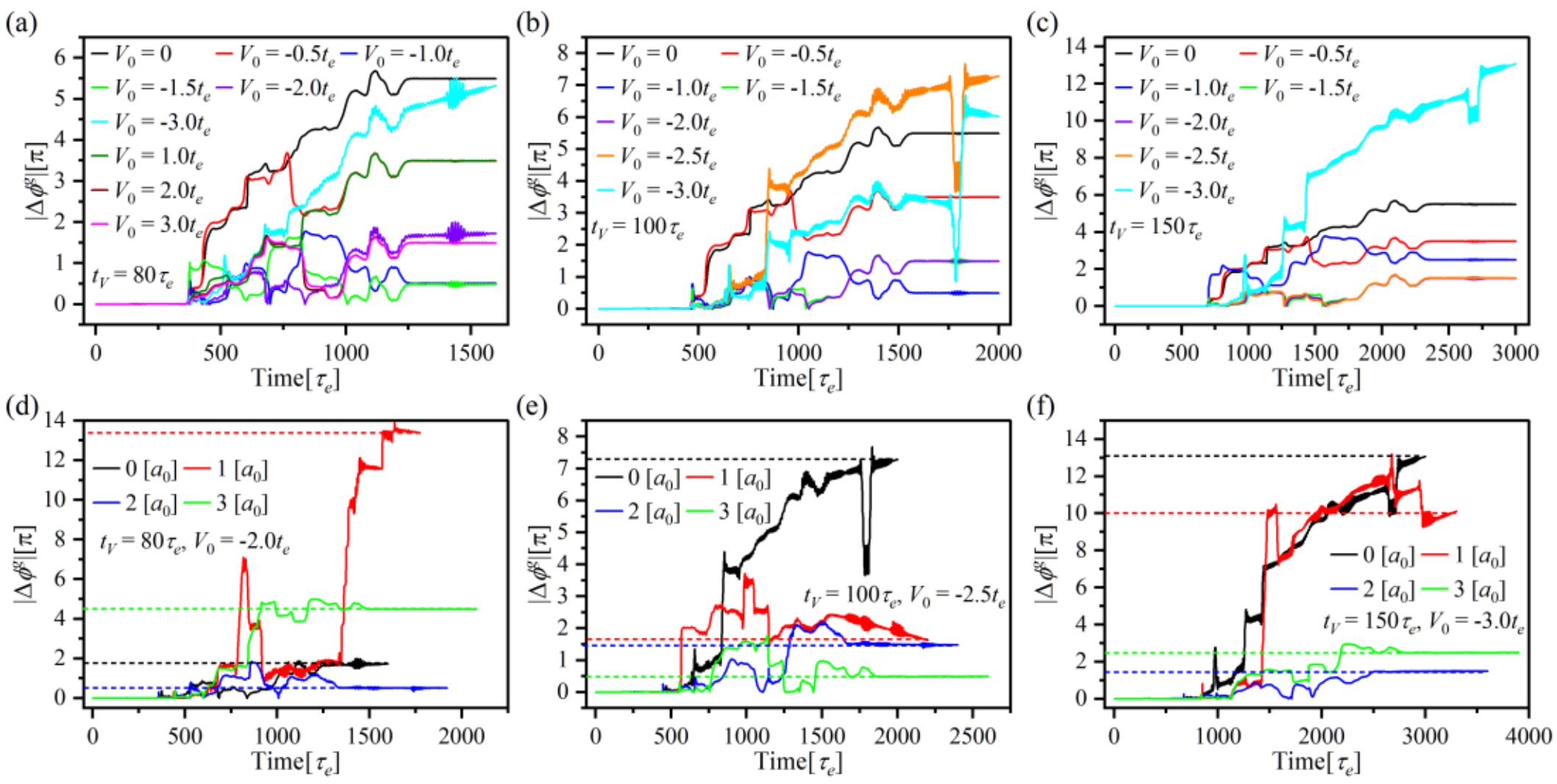

FIG. 3. The absolute value of geometric phase difference $\left|\Delta\phi^g\right|$ accumulated during the braiding process. (a)–(c) Braiding along the path of Fig. 1 for (a) $V_0 \in [-3.0, 3.0]$ at $t_V = 80\tau_e$, (b) $V_0 \in [-3.0, 0]t_e$ at $t_V = 100\tau_e$, and (c) $V_0 \in [-3.0, 0]$ at $t_V = 150\tau_e$. (d)–(f) Braiding along the path of Fig. 7 with a minimum vortex–impurity distance of 0, 1, 2, and 3 [$a_0$], (d) $t_V = 80\tau_e$, $V_0$ = -2.0$t_e$, (e) $t_V = 100\tau_e$, $V_0$ = -2.5$t_e$, (f) $t_V = 150\tau_e$, $V_0$ = -3.0$t_e$.

The accumulated geometric-phase difference provides a sensitive diagnostic of impurity-induced deviations from geometric-phase quantization as show in Fig. 3. For Ising exchange statistics, the phase difference upon completion of the braid should be an odd multiple of $\pi/2$, modulo $2\pi$. Figs. 3 (a)–(c) correspond to the direct path of Fig. 1. For repulsive impurities at $t_V = 80\tau_e$, the completed calculations at $V_0 = +1, +2, +3t_e$ yield $\left|\Delta\phi^g\right|/\pi$ = 3.491, 1.493, 1.492, respectively, all close to quantized values. For attractive impurities at the same $t_V$, near-quantization persists through $V_0 = -1.5t_e$, where $\left|\Delta\phi^g\right|/\pi$ = 0.482, only 0.018 below 0.5. Clear deviations occur at $V_0 = -2t_e$ and $-3t_e$ , for which the corresponding values are 1.725 and 5.321. Slowing the vortex motion extends the range of impurity strengths retaining near-quantization: the largest calculated strengths are $V_0$ = -1.5, -2.0, -2.5$t_e$ at $t_V$ = 80, 100, 150$\tau_e$, respectively. In particular, at $t_V = 150\tau_e$, the values for $V_0 = -2t_e$ and $-2.5t_e$ return to 1.503 and 1.524. This trend is consistent with the gap-bottleneck picture in Fig. 2 and with improved adiabaticity at lower vortex speeds.

Increasing the vortex–impurity clearance along the detoured paths in Fig. 7 systematically reduces the geometric-phase deviations [Figs. 3(d)–3(f)], calculated for ($t_V$, $V_0$) = ($80\tau_e$, $-2t_e$), ($100\tau_e$, $-2.5t_e$), and ($150\tau_e$, $-3t_e$), respectively. In all three cases, a clearance of 1[$a_0$] remains insufficient to restore quantization, with phase deviations of 0.088–0.393π. These deviations decrease to 0.013–0.022π at 2[$a_0$] and 0.009–0.018π at 3[$a_0$]. Thus, 2[$a_0$] essentially restores phase quantization, while 3[$a_0$] maintains comparable phase accuracy; the further suppression of the $N_{\mathrm{neq}}$ distortion at 3[$a_0$] in Fig. 7 is consistent with a reduced vortex–impurity overlap.

Notably, the visually stronger distortion induced by $V_0 > 0$ leaves the geometric-phase quantization intact, while $V_0 < 0$, despite its milder visual signature, destroys the $\sqrt{Z}$ gate by suppressing the local topological gap: the $N_{\text{neq}}$ texture distortion and the gate damage are two distinct effects of the impurity.

## B. Nonmagnetic impurity effects on $\sqrt{X}$ and $X$ Construction

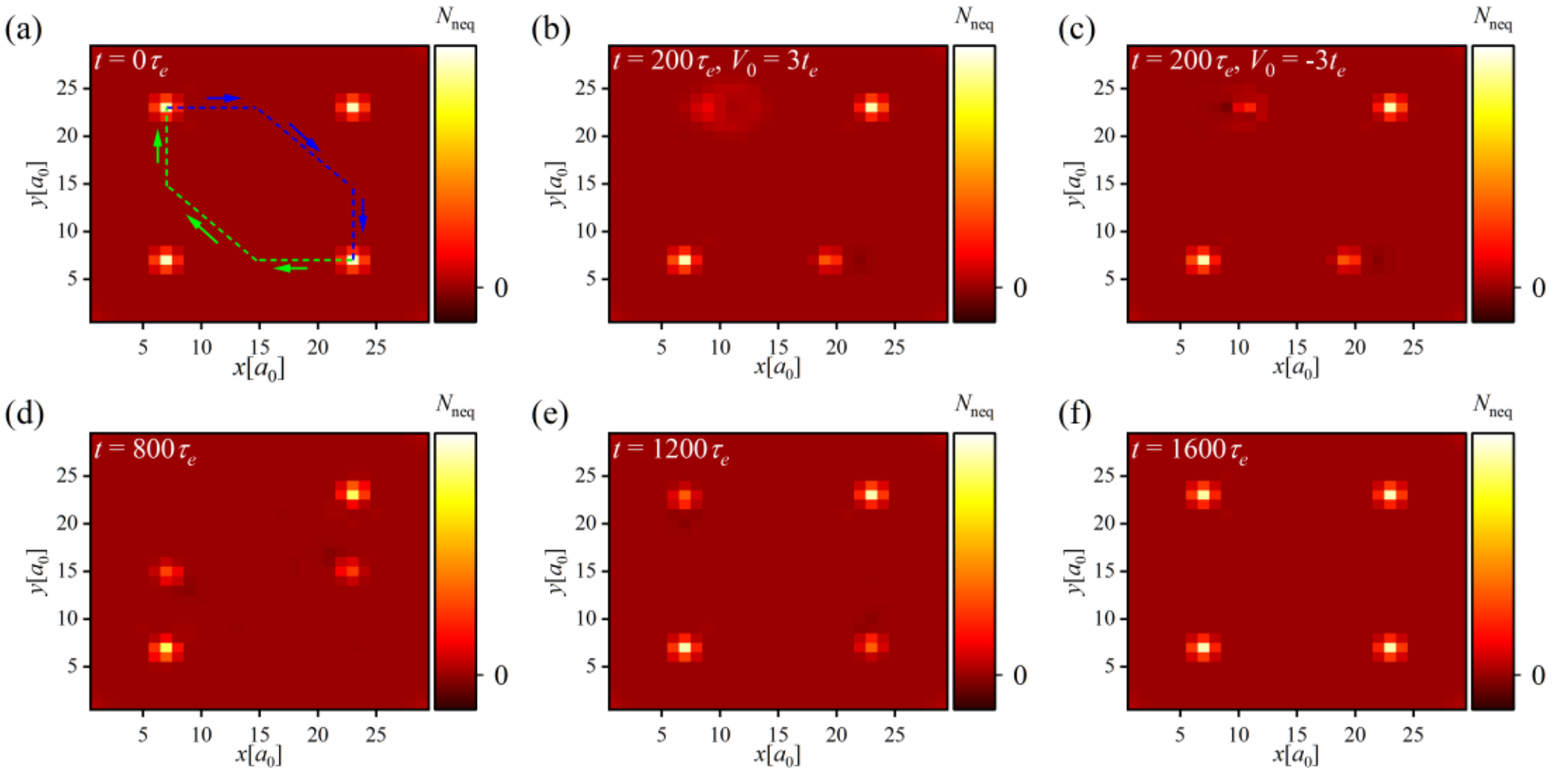

FIG. 4. Nonequilibrium zero-energy LDOS $N_{\text{neq}}(t, \omega = 0, r)$ visualizing the braiding of two vortex-core MZMs for $\sqrt{X}$ gate at (a) $t = 0\tau_e$, (b) $t = 200\tau_e$, $V_0 = 3t_e$, (c) $t = 200\tau_e$, $V_0 = -3t_e$, (d) $t = 800\tau_e$, (e) $t = 1200\tau_e$, and (f) $t = 1600\tau_e$. The blue (green) arrow indicates the direction of motion of the two moving vortices initially located at (7, 23), (23, 7); dashed lines mark the vortex trajectories. Distances are in units of the lattice constant $a_0$. $t_v = 50\tau_e$.

The evolution of $N_{\text{neq}}$ reveals how a nonmagnetic impurity redistributes the zero-energy weight during braiding in the four-vortex configuration as show in Fig. 4. Since the spatial distributions for the two impurity signs are very similar at the times shown in Figs. 4(a) and 4(d)–4(f), these panels display only representative results for $V_0 = +3t_e$. Figs 4(b) and 4(c) compare the repulsive and attractive impurities at $t = 200\tau_e$. As shown in Fig. 4(a), zero-energy peaks with heights of approximately 6.3–6.4 initially appear at the four vortex cores: the moving vortices at (7,23) and (23,7), and the static vortices at (7,7) and (23,23). The impurity site is (11, 23). At t=$200\tau_e$, one moving vortex crosses the impurity. In Fig. 4(b), the repulsive impurity suppresses the central weight to 0.1386, producing a local minimum surrounded by a ring-like distribution. In Fig. 4(c), the attractive impurity leaves a central local maximum of 1.4508, accompanied by a negative side lobe of -0.3301 at (9,23). These features are qualitatively similar to those in the two-vortex case of Fig. 1. At this time, the peaks of the distant moving vortex and the two static vortices are nearly identical for the two impurity signs. After the crossing vortex moves away, the differences between the two impurity signs decrease rapidly at $t = 800\tau_e$ and $1200\tau_e$, respectively. The positive-impurity results in Figs. 4(d), 4(e) and 4(f) are therefore representative of both signs.

The vortex paths are completed at $t = 1200\tau_e$, and Fig. 4(f) shows the subsequent state at t = $1600\tau_e$. For $V_0 = +3t_e$, the two moving-vortex peaks then have heights of 6.211 and 6.330, while the static-vortex peaks remain approximately 6.40.

On the basis of Fig. 4(e), the two moving vortices in the four-vortex configuration continue to braid along the dashed trajectories shown in Fig. 4(a); the resulting complete double braid realizes the $X$ gate.

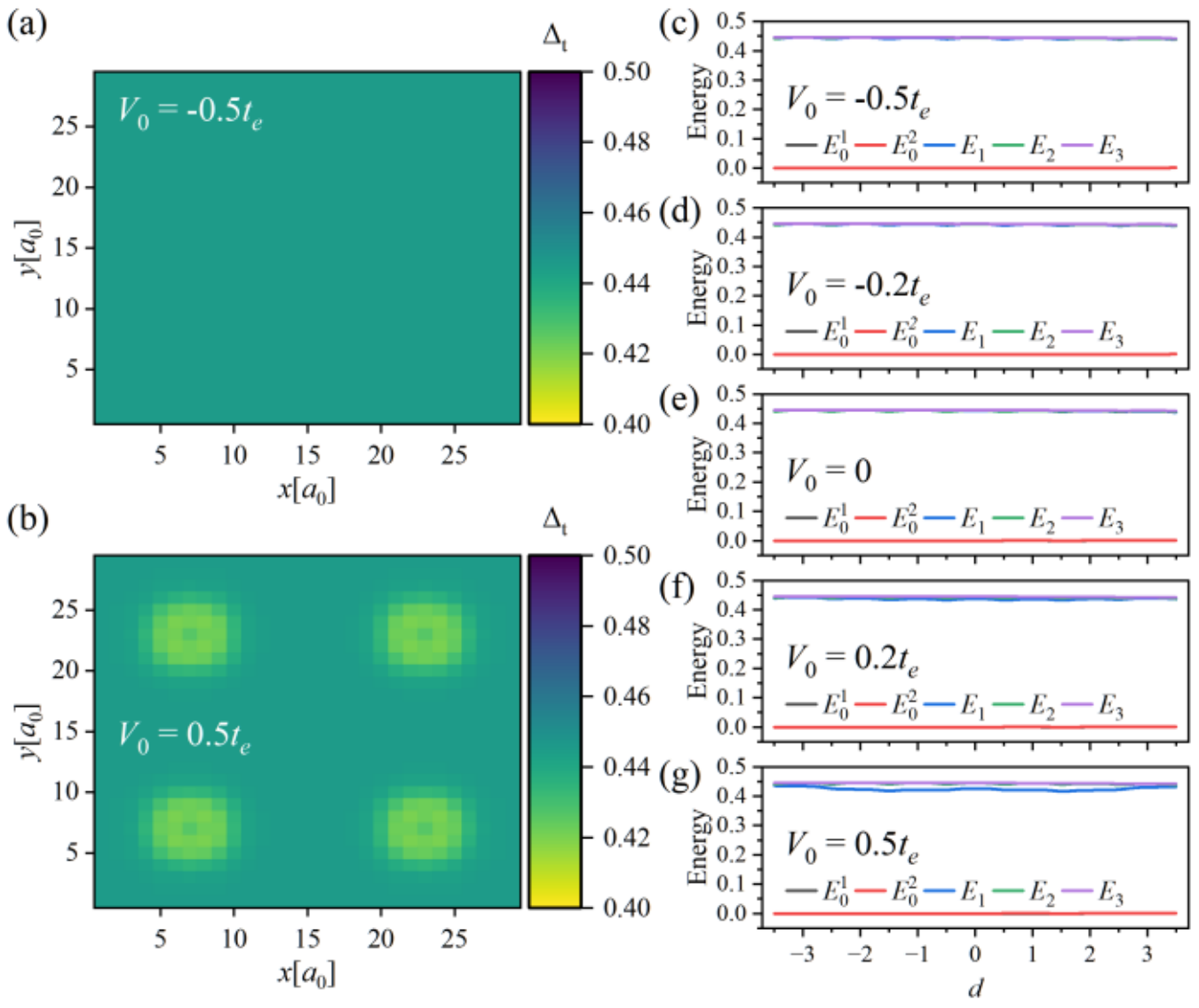

FIG. 5. Effect of a nonmagnetic impurity on the near-zero-energy spectrum of four-vortex configuration. (a), (b) Topological gap $\Delta_t$ as a function of the impurity position for (a) $V_0 = -0.5t_e$ and (b) $V_0 = 0.5t_e$; (c)–(g) Energies of the two MZMs and of the first, second, and third excited states as a function of the vortex–impurity distance $d$ for (c) $V_0 = -0.5t_e$, (d) $V_0 = -0.2t_e$, (e) $V_0 = 0$, (f) $V_0 = 0.2t_e$, and (g) $V_0 = 0.5t_e$. Energies are in units of $t_e$. Four vortices initially located at (7, 7), (7, 23), (23, 7), (23, 23).

The four-vortex configuration exhibits a different impurity-sign dependence of the topological gap as show in Fig. 5. For $V_0 = -0.5t_e$, $\Delta_t$ varies only approximately $0.445t_e$ over the entire impurity-position scan. In contrast, for $V_0 = 0.5t_e$, localized gap-suppression regions appear around the four vortices, with a minimum value of approximately $0.420t_e$, whereas the gap recovers to approximately $0.445t_e$ away from the vortices. The two positive-energy MZM levels remain below $10^{-3}t_e$ throughout the plotted distance range. The spectra for attractive impurities ($V_0 = -0.2t_e$ and $-0.5t_e$) nearly coincide with that of the clean system, whereas repulsive impurities mainly suppress the first excited state, with the suppression becoming stronger as $V_0$ increases. At $d = 0$, $E_1$ decreases from $0.4446t_e$ in the clean system to $0.4375t_e$ for $V_0 = 0.2t_e$ and $0.4260t_e$ for $V_0 = 0.5t_e$.

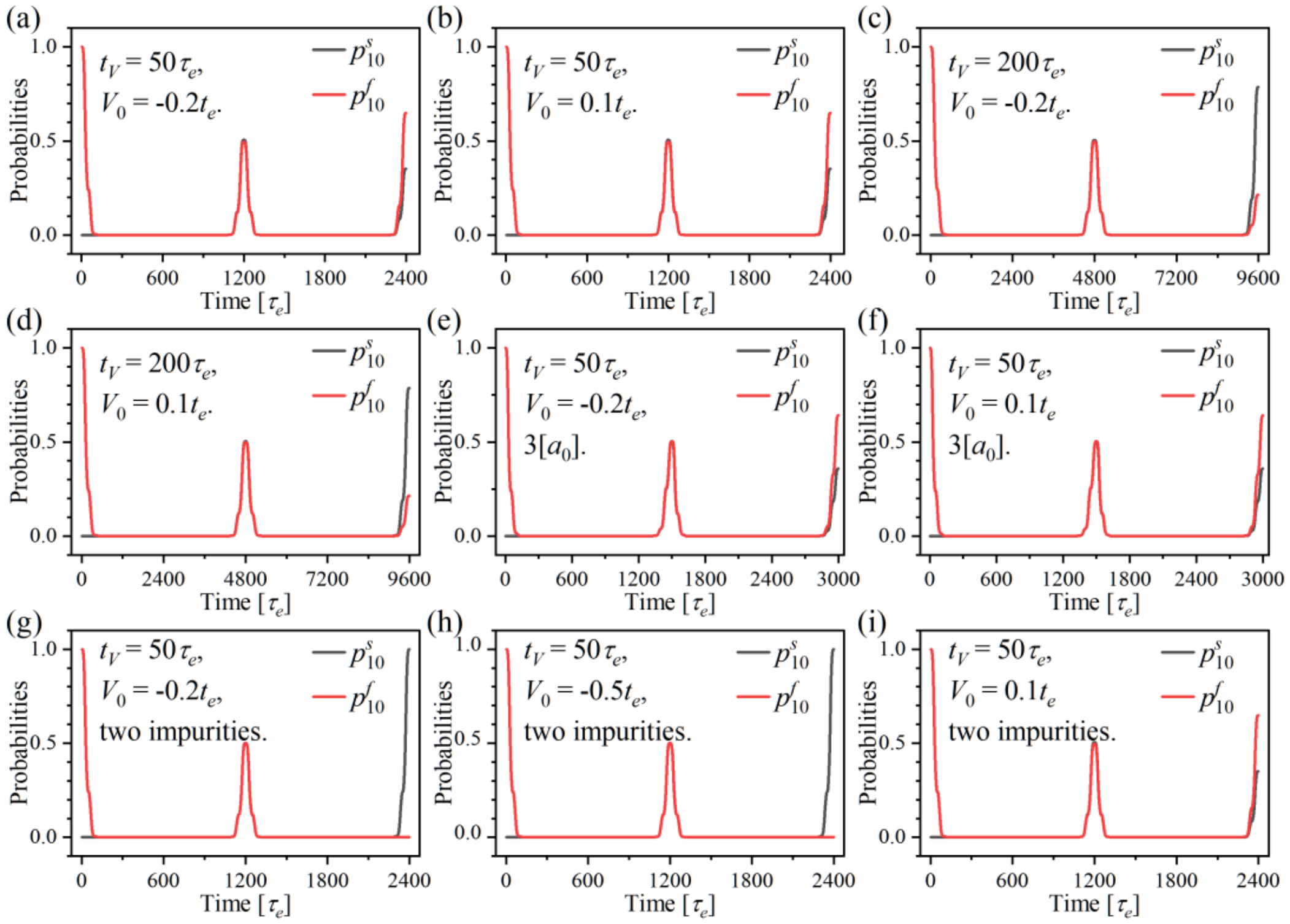

FIG. 6. Time evolution of the probabilities $p^f_{10}$ (red curves) and $p^s_{10}$ (black curves). (a)–(d) follow the braiding paths shown in Fig. 4 with a single impurity located at (11,23): $t_v = 50\tau_e$ in (a) and (b), and $t_v = 200\tau_e$ in (c) and (d); $V_0 = -0.2t_e$ in (a) and (c), and $V_0 = 0.1t_e$ in (b) and (d). (e) and (f) follow the detoured paths shown in Fig. 8, with a vortex–impurity clearance of 3[$a_0$] and a single impurity at (11,23); $t_v = 50\tau_e$ , with $V_0 = -0.2t_e$ in (e) and $V_0 = 0.1t_e$ in (f). (g)–(i) follow the paths shown in Fig. 4 with two equal-strength impurities located at (11,23) and (19,7); $t_v = 50\tau_e$, and $V_0 = -0.2t_e$, $-0.5t_e$ , $0.1t_e$ in (g), (h), and (i), respectively.

The first exchange produces an approximately equal-weight superposition, with $p^f_{10} \simeq p^s_{10} \simeq 0.5$ in all nine cases, whereas the main differences emerge after the second exchange as show in Fig. 6. For a single impurity on the paths shown in Fig. 4, the final success probabilities at $t_v = 50\tau_e$ are 0.3512 for $V_0 = -0.2t_e$ and 0.3514 for $V_0 = 0.1t_e$. Increasing $t_v$ to $200\tau_e$ raises these values to 0.7864 and 0.7859, respectively, consistent with reduced nonadiabatic errors during slower braiding. At $t_v = 50\tau_e$, the 3[$a_0$] detour yields only modest increases to 0.3584 and 0.3586, respectively, indicating that the detour provides little improvement despite the increased vortex–impurity clearance. In contrast, two equal-strength impurities with $V_0 = -0.2t_e$ and $-0.5t_e$ produce final success probabilities of 0.9986 and 0.9985, respectively, with failure probabilities below $8\times10^{-4}$, whereas two $+0.1t_e$ impurities yield a success probability of only 0.3515. Since $p^f_{10} + p^s_{10} > 0.9992$ at the end of every protocol, these differences primarily reflect the redistribution of occupation between the two computational states during the coherent evolution after the second exchange, rather than appreciable leakage out of the computational subspace. In particular, the near-unit success probability obtained with two negative impurities constitutes a pronounced two-impurity effect that cannot be inferred from the single-impurity results.

## IV Conclusion and Future Prospects

We have investigated two- and four-vortex Majorana braiding gates using a time-dependent BdG model with nonmagnetic Gaussian impurities, complemented by geometric-phase, many-body-overlap, and nonequilibrium LDOS diagnostics. In the two-vortex configuration, sufficiently strong attractive impurities suppress the $\Delta_t$, thereby reducing the adiabaticity of the braiding process and degrading the quantization of the geometric phase difference, whereas the repulsive impurities considered here cause no comparable gate damage despite producing stronger zero-energy LDOS distortions. Slower vortex motion extends the range of impurity strengths retaining near-quantization, and a 2[$a_0$] detour essentially restores quantization for the parameters examined. The four-vortex configuration exhibits a different sign dependence: attractive impurities leave the spectrum close to the clean-system result, whereas repulsive impurities mainly suppress the first excited state. A longer braiding time substantially improves the $X$ gate success probability, while a 3[$a_0$] detour provides little improvement for fast braiding. Notably, a specific pair of attractive impurities increases the success probability from approximately 0.35 for the corresponding single-impurity case to approximately 0.9985, demonstrating a pronounced two-impurity effect. These results establish that LDOS distortions are not equivalent to gate damage and that impurity effects must be evaluated through the configuration-dependent spectrum and full braiding dynamics. Future work should clarify the microscopic origin of the enhanced success probability for paired attractive impurities, examine its robustness against impurity-parameter variations and multi-impurity disorder, and jointly optimize braiding paths and speeds to assess the potential of impurity engineering for reliable Majorana gates.

**Acknowledgment** This work is supported by the National Natural Science Foundation of China (Grant No. 11204261), College of Physics and Optoelectronic Engineering training program (Grant No. 10QDZ18), a Key Project of the Education Department of Hunan Province (Grant No. 19A471), Natural Science Foundation of Hunan Province (Grant No. 2018JJ2381), Natural Science Foundation of Hunan Province (Grant No. 2026JJ90084).

**Data Availability Statement** The data that support the findings of this article are openly available [47].

**Declarations**

**Conflict of interest** No potential conflict of interest was reported by the authors. All authors of this manuscript have read and approved the final version submitted, and contents of this manuscript have not been copyrighted or published previously and are not under consideration for publication elsewhere.

**ORCID iDs**
Jiangwei Long 0009-0000-5179-0370
Zihui Liu 0009-0002-1064-9340
Yizhi Li 0000-0001-8056-2684
Lijun Meng 0000-0003-4147-6411

## Appendix A: $N_{\mathrm{neq}}$ visualization of the detoured braiding paths

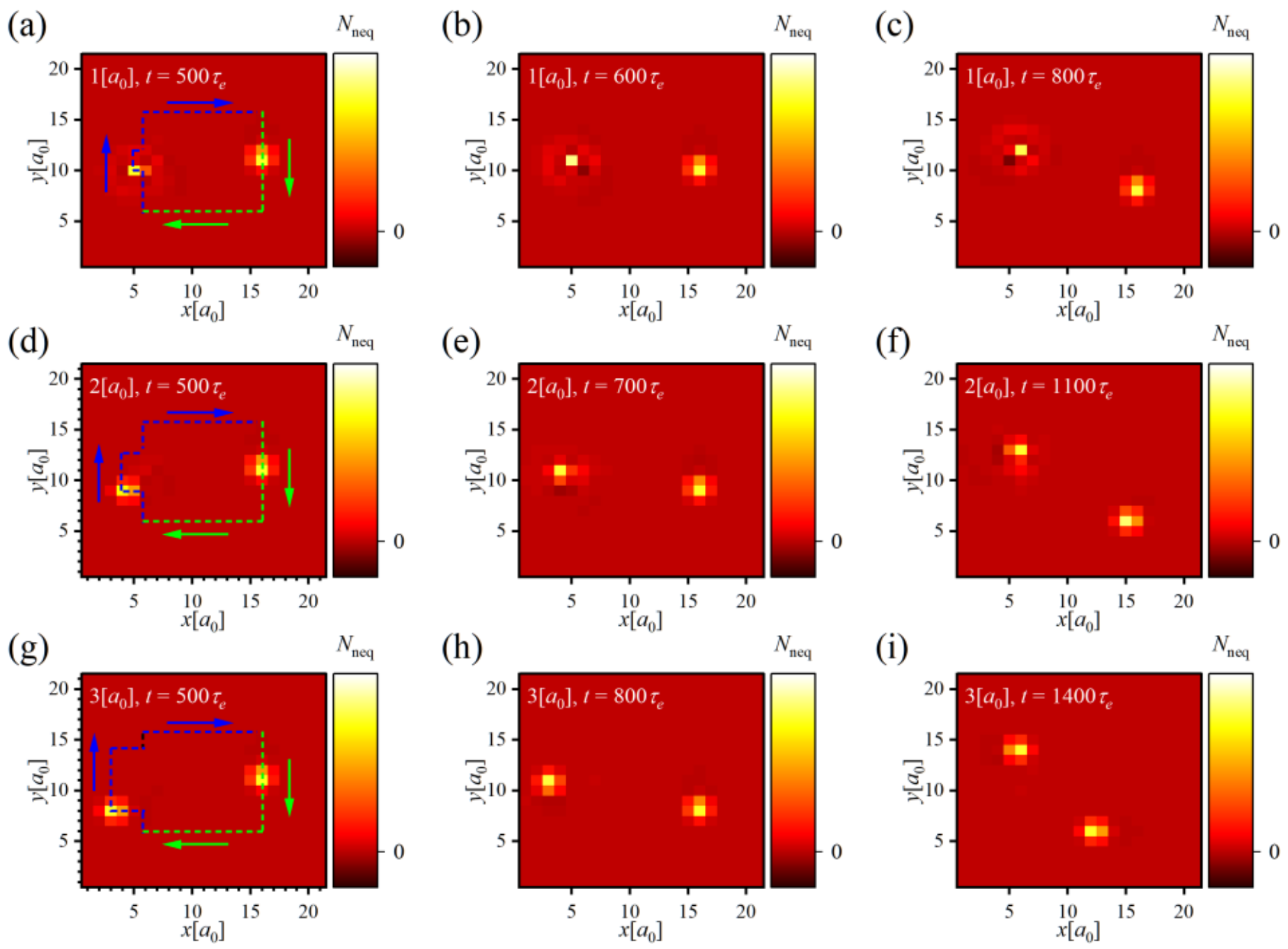

FIG. 7. Nonequilibrium zero-energy LDOS $N_{\mathrm{neq}}(t, \omega = 0, r)$ along the detoured braiding paths for two-vortex configuration. The impurity is located at (6, 11) ($V_0 = -3$, η = 1), and the minimum vortex–impurity distance is (a)–(c) 1[$a_0$] ($t$ = 500, 600, 800$\tau_e$), (d)–(f) 2[$a_0$] ($t$ = 500, 700, 1100$\tau_e$), and (g)–(i) 3[$a_0$] (t = 500, 800, 1400$\tau_e$). The blue (green) arrow indicates the direction of motion of the two vortices initially located at (6, 6), (16, 16); dashed lines mark the vortex trajectories. $t_V$ = 100$\tau_e$.

The zero-energy LDOS response along the three detoured paths used in Figs. 3(d)–3(f) depends strongly on the vortex–impurity clearance as show in Fig. 7. In contrast to direct passage through the impurity [Fig. 1(c)], which spatially contracts the zero-energy peak and produces a pronounced adjacent negative lobe, increasing the detour clearance strongly suppresses the impurity-induced zero-energy LDOS distortion. Using the strongest negative lobe among the three snapshots for each path, the minimum reaches -0.45 at 1[$a_0$], nearly an order of magnitude larger in magnitude than the clean background, while the moving-vortex zero-energy peak remains pronounced with heights of 3.6–4.3. At 2[$a_0$], the strongest negative lobe is reduced to -0.19. At 3[$a_0$] , it is only approximately -0.06, close to the clean negative background of about -0.05, and the moving-vortex peak heights range from 3.7 to 3.8. This suppression of the real-space distortion is consistent with the recovery of geometric-phase quantization from 2[$a_0$] onward in Fig. 3(d)–(f). The $N_{\mathrm{neq}}$ maps thus provide direct real-space evidence for the suppression of impurity-induced zero-energy LDOS distortions and, together with the independent geometric-phase and fidelity results, support the restoration of the gate along the detoured paths.

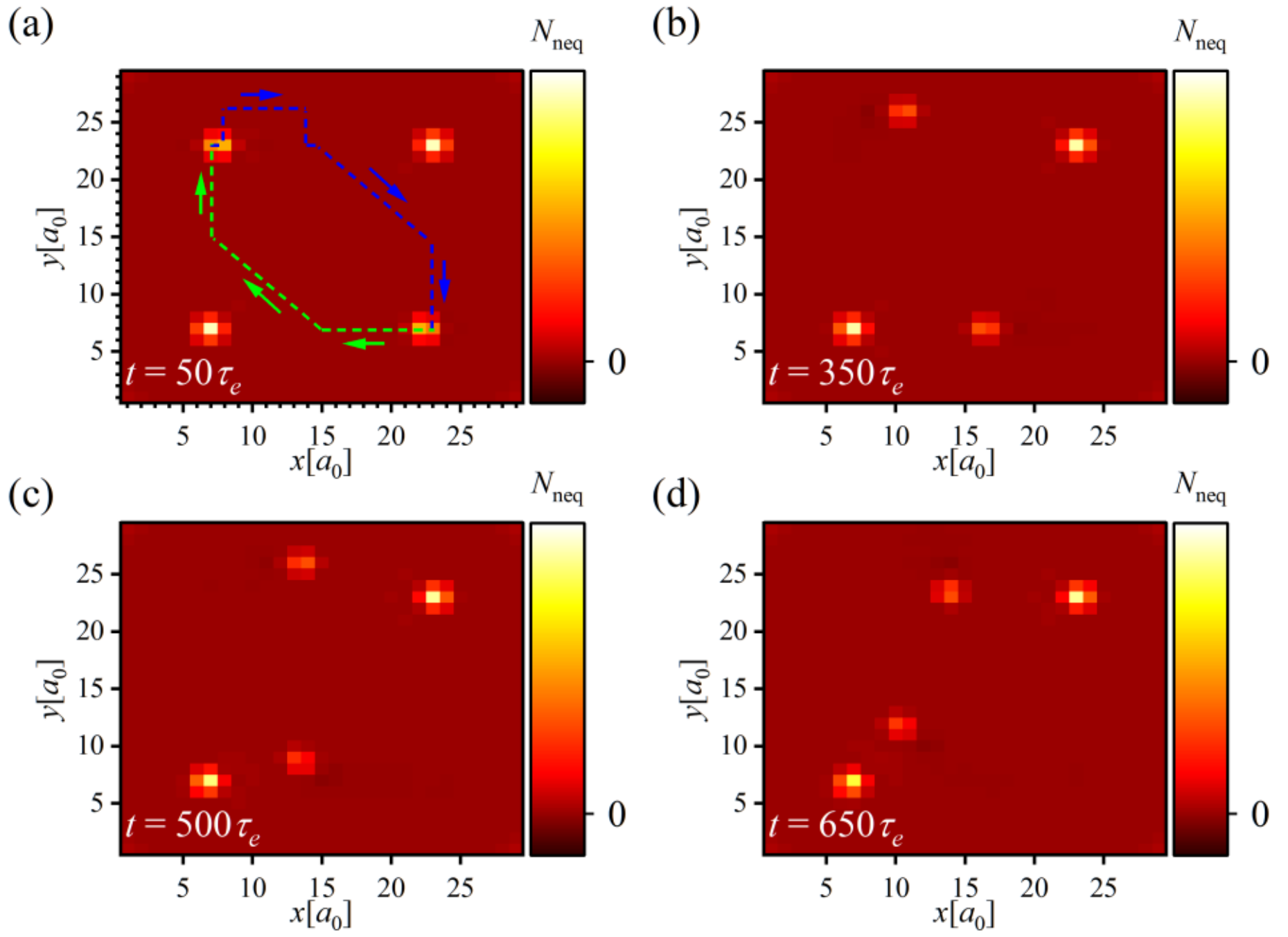

FIG. 8. Nonequilibrium zero-energy LDOS $N_{\mathrm{neq}}(t, \omega = 0, r)$ along the detoured braiding paths for four-vortex configuration. The impurity is located at (11, 23) ($V_0 = -3$, η = 1), and the minimum vortex–impurity distance is 3[$a_0$] (a) $t = 50\tau_e$, (b) $t = 350\tau_e$, (c) $t = 500\tau_e$, (d) t = $650\tau_e$. The blue (green) arrow indicates the direction of motion of the two vortices initially located at (7, 23), (23, 7); dashed lines mark the vortex trajectories. $t_V = 50\tau_e$.

Along the detoured paths in Figs. 6(e) and 6(f), $N_{\mathrm{neq}}$ tracks the four-vortex Majorana zero modes and reveals the residual real-space influence of the impurity as show in Fig. 8. A nonmagnetic impurity with strength $V_0 = -3$ and spatial extent η = 1 is located at (11,23), and the detour maintains a minimum vortex–impurity separation of $3a_0$. At all the selected times, the zero-energy LDOS maxima associated with the two moving vortices remain clearly resolved and follow the vortex cores along the dashed trajectories, whereas those associated with the two stationary vortices remain at their original positions. In particular, no pronounced impurity-induced zero-energy peak emerges near the impurity site, and no appreciable distortion or loss of localization of the moving maxima is visible as the vortices pass through the impurity region. Thus, at the level of the real-space $N_{\mathrm{neq}}$ distribution, the visual influence of the impurity on the motion of the vortex-core zero modes appears nearly negligible after introducing the detour. It should nevertheless be emphasized that the LDOS does not contain the full dynamical phase and coherence information required to determine the success probability of a braiding gate. Within the braiding convention adopted here, a single exchange along the path shown in Fig. 8(a) implements the $\sqrt{X}$ gate, while a second successive exchange along the same geometric path is required to complete the full $X$ gate. Whether this detoured protocol restores the success probability of the $X$ gate must therefore be established by independent time-dependent many-body evolution and overlap calculations, rather than inferred from the LDOS snapshots alone.